\documentstyle[12pt,epsfig]{article}
\newcommand{\beq}{\begin{equation}}
\newcommand{\eeq}{\end{equation}}
\newcommand{\beqar}{\begin{eqnarray}}
\newcommand{\eeqar}{\end{eqnarray}}
\newcommand{\noi}{\noindent}
\newcommand{\bra}[1]{\langle #1|}

\def\kket#1{| #1\rangle}

\newcommand{\sect}[1]{\section{#1}}
\newcommand{\ssect}[1]{\subsection{#1}}
\newcommand{\lsim}{\mathrel{\rlap{\lower4pt\hbox{\hskip0pt$\sim$}}
\raise1pt\hbox{$<$}}}
\newcommand{\gsim}{\mathrel{\rlap{\lower4pt\hbox{\hskip0pt$\sim$}}
\raise1pt\hbox{$>$}}}

\begin{document}
\noindent
\hspace*{12.5cm}June. 1997\\
\hspace*{12.5cm}TOYAMA-96\\

\vspace*{2cm}

\begin{center}
  {\huge Toward the second stage at B factory\footnote{%
To appear in the proceedings of ''Mass and Mixing of Quarks and Leptons" 
at Shizuoka, 1997}
\\ \vspace{3mm}
}

\end{center}
\vspace*{0.1cm}
\begin{center}
{\Large T.~Kurimoto}\footnote{e-mail: krmt@sci.toyama-u.ac.jp}\\
\vspace*{.2cm}
\begin{large}
  
Department of Physics, Faculty of Science,\\
Toyama University,\\
Toyama 930, Japan\\

\vspace*{3cm}
\Large{\bf Abstract} 
\end{large}
\end{center}

The measurement and precision at B physics experiment are reviewed 
by taking into account of the numbers of B mesons to be produced in 
the future experimental projects. With $10^9$ or more B mesons we 
will be able to fix the parameters in Kobayashi-Maskawa matrix elements
or find a signal of new physics.

\newpage
\sect{Measurements at B factories}
The gauge interaction in the standard model (SM) has been well understood 
and analyzed through the experiments so far made. 
The coming experiments 
are aimed for the detailed research of Yukawa interaction and Higgs interaction  
to fully understand SM and explore new physics beyond it. The quark flavor 
mixing matrix proposed by Kobayashi and Maskawa (KM) \cite{KM} is closely 
related to Yukawa and Higgs sector. Precise determination of the 
KM matrix elements gives us valuable insight about SM and new physics. 
The physics of hadrons containing $b$ quark, so called B physics,  
is indispensable for the determination of KM matrix elements concerning with 
the third generation quarks. So B physics is taken up as one of the main 
projects in many of the accelerator experiments to be done in the near future:  
Two anti-symmetric $e^+e^-$ colliders dedicated for B physics are 
now under construction at KEK \cite{KEKB} and SLAC \cite{SLACB}. 
A dedicated experiment is planned by using the proton beam of the    
$ep$ collider HERA at DESY \cite{HERAB}. These new experiments will begin 
physics run in 1999. 
CLEO group has been working on B physics and reported valuable results  
by using the symmetric $e^+e^-$ collider CESR at Cornell \cite{CLEO}. The 
CESR will be upgraded for more luminosity and do physics run (CESR phase-III). 
The Tevatron $p\bar p$ collider has also been giving information on 
B physics and will be upgraded with main injector \cite{TEVA}. 
B physics will be explored also at the high-energy high-luminosity $pp$ collider 
LHC now under construction at CERN \cite{LHCB}. Above is summarized in Table 1 with 
the year and possible number of B meson produced.
\begin{center}
\begin{tabular}[c]{|c|c|c|} \hline
year & \# of {\bf B}& facilities \\ \hline \hline
$\leq  2000$ & $10^{7\sim 8}$\rule{0mm}{6mm} &  Tevatron, CESR phase-III, HERA-B,\\
& &  KEK-B, SLAC-B\\ \hline
$\gsim 2010$ & $\gsim 10^9 $ \rule{0mm}{6mm}&  + LHC-B, Tevatron (Main Injector) \\ \hline
\end{tabular}
\vspace*{4mm}

Table 1 : Experimental facilities of B physics and number of B mesons to be produced. 

\end{center}

\noi
The B physics experiments can be divided into two stages. We get $10^{7\sim 8}$  B 
mesons at the first stage within this century, where the main goal is 
to obtain the first evidence of CP violation in B meson system. It is promising 
because of the existence of the so called golden mode,
 $B^0, \overline{B^0} \rightarrow J/\Psi K_S$, which has relatively large 
branching ratio ($O (10^{-4})$) and can be identified by clean signal; 
$J/\Psi \rightarrow l^+l^-$, $K_S \rightarrow \pi^+\pi^-$. A decade after the 
beginning of the 21st century we will obtain more than $10^9$ of B, and then 
we will be able to fix the KM matrix elements more precisely which enables us 
to explore new physics beyond SM. This is the second stage. 
In this paper we summarize the precision of KM matrix determination when experiments 
are done well as proposed, and see how new physics search can be made. 

To begin with, let us review briefly what are measured in B physics experiments. 
KM matrix $V$ appears in the interaction among quark charged 
currents and $W$ boson
\beq
{\cal L}_W = \frac{g}{\sqrt{2}}\left[
\overline {u_{Li}}\gamma^\mu (V)_{ij}d_{Lj}W^+_\mu 
  + \overline {d_{Lj}}\gamma^\mu (V^*)_{ij}u_{Li}W^-_\mu \right].
\eeq
The matrix $V$ is a $3\times3$ unitarity matrix in the three generation 
standard model, so that 
the following condition holds :
\beq
{V_{ub}}^* V_{ud} + {V_{cb}}^* V_{cd} + {V_{tb}}^* V_{td} =0.
\eeq  
We have so-called unitarity triangle by expressing the above condition 
in complex plane.
\begin{figure}[hbtp]
  \begin{center}
    \leavevmode
    \epsfig{figure=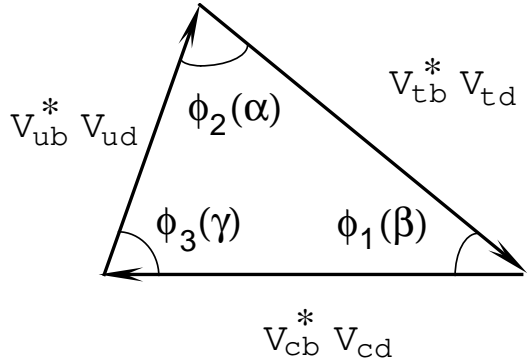,width=7cm}
    \caption{Unitarity triangle}
    \label{fig:unitri}
  \end{center}
\end{figure}
The parameters in KM matrix can be determined 
through the measurements of the sides and the angles of this triangle. 
The KM matrix elements $|V_{ud}|$ and $|V_{cd}|$ have already been measured 
to 0.1 \% and 7 \%, respectively \cite{PDG}. The elements $|V_{cb}|$ and $|V_{ub}|$ 
can be obtained through the semi-leptonic decays of B meson, 
$ b \rightarrow cl\bar\nu$ and $ b \rightarrow ul\bar\nu$. 
The magnitude of $B^0$-$\overline {B^0}$ mixing gives the side $|{V_{tb}}^* V_{td}|$ 
because the top quark contribution dominates in the the box diagram \cite{ilh}. 
The length of this side can be calculated by utilizing three generation unitarity 
also from the ratio $|V_{td}/V_{ts}|$ which 
can be measured through 
$\Gamma (b\rightarrow d\gamma )/\Gamma (b\rightarrow s\gamma )$ or 
the ratio of $B^0$-$\overline {B^0}$ mixing to 
$B^0_s$-$\overline {B^0_s}$ mixing. 
The angles can be obtained through the CP violation measurements:  
CP violation in the mode $B^0, \overline{B^0} \rightarrow J/\Psi K_S$ gives 
 $\sin 2\phi_1$.\footnote{
There are two notations for the angles of unitarity triangle. 
The one used here is $\phi_{1,2,3}$ and another ($\alpha$, $\beta$ and $\gamma$)
is taken in ref.\cite{PDG}. I prefer the former notation as 
$\alpha$, $\beta$ and $\gamma$ are often used to express another quantities.
}
The mode $B^0, \overline{B^0} \rightarrow \pi\pi$ or $\rho\pi$ gives $\sin 2\phi_2$ 
in the same way \cite{cpang}. The rest of the angle $\phi_3$ can be 
measured through the direct CP violation 
in $B^\pm \rightarrow (D^0,\overline{D^0}, D_{CP})K^\pm$ \cite{phit}. It 
can also be obtained in $B^0_s, \overline{B^0_s} \rightarrow \rho K_S$ decay 
when $B^0_s$ mason is available \cite{cpang}. 
\sect{Precision of measurements}
Here  we summarize the present status of these measurements \cite{brow} 
and the precision reach in the coming experiments. 
The precision is estimated based on the simulation results given in 
1996 BELLE progress reports \cite{bprep} by assuming everything goes well 
in experimental side.

\ssect{$|V_{cb}|$}
$|V_{cb}|$ is obtained through semi-leptonic $b\rightarrow c$ decay in both 
inclusive and exclusive modes. Determination of $|V_{cb}|$ by using inclusive 
mode suffers from relatively large ($\sim $10 \%) theoretical uncertainty; 
calculation of higher order QCD correction, estimation of non-perturbative 
effects, values of $b$ quark mass and so on. Heavy quark symmetry allows us 
less uncertain determination of $|V_{cb}|$ in exclusive mode \cite{hqet}.
The value of form factor at zero-recoil limit can be predicted from the 
symmetry. With more statistics exclusive mode should 
become promising for the precise 
determination of $|V_{cb}|$. Present data from exclusive $B \rightarrow D^* l\nu$ 
decay is as follows \cite{fort};
\beq
|V_{cb}| = (34.3 \pm 2.4 \pm 1.3) \times 10^{-3},  
\eeq 
where the first error is experimental and the second due to theoretical 
uncertainty \cite{neub}. 
We now have about 7 \% experimental error and 3 \% theoretical error concerning on 
QCD correction and finite mass correction. 
The statistical error can be reduced to 0.3 \% with $10^8$ B meson 
due to simulation. Then assuming the systematic error is 
same order as the statistical error, the experimental error 
will be reduced to about 1 \% or less. The experimental error will become negligible 
in comparison with the theoretical error with more than $10^9$ B in the next century 
unless there emerges drastic theoretical improvement.
\ssect{$|V_{ub}|$}
$|V_{cb}|$ is obtained in a similar manner as $|V_{cb}|$ by using 
$b\rightarrow u$ semi-leptonic decay. The high energy lepton at the 
lepton energy spectrum is used to identify $b\rightarrow u$ transition 
in inclusive mode. CLEO has succeeded in identifying exclusive mode;
$B \rightarrow \pi l\nu$, $\rho l\nu$ \cite{buex}, and got
\beq
|V_{ub}| = (3.3 \pm 0.2 {\scriptsize
\begin{array}[c]{ll}
+0.3 \\ -0.4 
\end{array}}
\pm 0.7 )\times 10^{-3}, 
\eeq
where the first and second error is experimental and the third theoretical 
mainly due to hadron model dependence. Inclusive mode measurement has given 
similar value ($|V_{ub}|=(3.2\pm0.8) \times 10^{-3}$). We have at present 
experimental and theoretical errors of both about 20 \%. Here heavy quark symmetry 
cannot be used to get a value of form factor. So large theoretical 
ambiguity cannot be avoided with present technique. Experimental error can be 
reduced to about 10 \% with statistics of $10^8$ B mesons. 
Theoretical error will be dominant also here in the future.       
\ssect{$|{V_{tb}}^*V_{td}|$}
Ten years has passed since the first discovery of $B^0$-$\overline {B^0}$ mixing.
The accumulated data now gives the mass difference between two mass eigenstates 
of neutral B mesons as  
\beq
\Delta M_B = 0.460 \pm 0.018 \mbox{ ps}^{-1},
\eeq
where the error is experimental only. This 4 \% error will be reduced 
to about 1 \% or less with more than $10^8$ B meson. The mass difference $\Delta M_B$ 
is related to $|{V_{tb}}^*V_{td}|$ in SM as follows;
\beq
\Delta M_B = 2 |\bra{B^0}H^{eff}\kket{\bar B^0}| 
\propto
|{V_{tb}}^*V_{td}|^2 B_B f_B^2, 
\eeq  
where $B_B f_B^2 M_B/3 \equiv 
\bra{B^0}(\overline{d_L}\gamma_\mu b_L)^2 \kket{\overline{B^0}}$.
The hadron matrix element $B_B f_B^2$ is from non-perturbative strong interaction 
which is hard to calculate precisely. Therefore the error in the determination 
of $|{V_{tb}}^*V_{td}|$ is dominated by theoretical uncertainty which is now 
about 20 \% in lattice QCD calculation. 

There are other methods to obtain the length of the side ${V_{tb}}^*V_{td}$ 
if we assume KM matrix is $3\times 3 $ unitary as in SM. Given the Wolfenstein 
parameterization of KM matrix \cite{wolf}, 
\beq
V= \left(\begin{array}{ccc}
  1-(\lambda^2/2) & \lambda & A\lambda^3(\rho-i\eta) \\
 -\lambda & 1-(\lambda^2/2) & A\lambda^2 \\
 A\lambda^3(1-\rho-i\eta) & -A\lambda^2 & 1
\end{array} \right), 
\eeq
we have 
\beqar
|{V_{tb}}^*V_{td}|/ |{V_{cb}}^*V_{cd}|&=&  \sqrt{(1-\rho)^2 +\eta^2}, \\
|{V_{td}}/V_{ts}| &=& \lambda \sqrt{(1-\rho)^2 +\eta^2}. 
\eeqar
The value of $\lambda =|V_{us}|$ is well known, so that $|{V_{td}}/V_{ts}|$ 
gives the length of the side. $|{V_{td}}/V_{ts}|$ can be obtained from 
the ratio of $B^0$-$\overline {B^0}$ mixing to  $B^0_s$-$\overline {B^0_s}$ mixing, or
the ratio of radiative penguin decays of $b$ quark.
\beqar
\Delta M_B /\Delta M_{Bs} &\propto & |{V_{td}}/V_{ts}|^2 ,\\
|A(b\rightarrow d \gamma )/ A(b\rightarrow s \gamma )|&\propto & |{V_{td}}/V_{ts}|.
\eeqar
Both $\Delta M_{Bs}$ and $b\rightarrow d \gamma$ decay have not yet measured, but 
will be obtained in the future experiments. One can expect 
about 20 \% experimental error with $10^8$  B mesons. Theoretical 
uncertainty lies in the evaluation of $SU(3)$ flavor symmetry breaking effect and 
long-distance effects. It depends on the theoretical development which of the 
measurements gives most precise value of the length of the side.
\ssect{$\phi_1$}
The angle $\phi_1$ is measured in the time dependent CP asymmetry in  
 $B^0, \overline{B^0} \rightarrow J/\Psi K_S$ decay.
\begin{eqnarray}
 Asy[f_{CP}] &\equiv& 
\frac{\Gamma[B^0 (t) \rightarrow f_{CP}] - 
      \Gamma[\overline{B^0}  (t)\rightarrow f_{CP}]
      }{
      \Gamma[B^0  (t)\rightarrow f_{CP}] + 
      \Gamma[\overline{B^0}  (t)\rightarrow f_{CP}] }\\
 &=&  \frac{2}{(2+c_d)} \left[\mbox{Im}(\frac{q}{p} \rho) \sin (\Delta M_B t)
                              -\frac{c_d}{2}\cos (\Delta M_B t)\right] ,
\end{eqnarray}
where $f_{CP}=J/\Psi K_S$ state here, and 
\begin{eqnarray}
  \frac{q}{p} \equiv \frac{|M_{12}^B|}{M_{12}^B},&\quad&
  M_{12}^B \equiv \langle B^0|{\cal H}^{\Delta B=2}| \overline {B^0}\rangle, \\
  \rho  \equiv  \frac{A(\overline{B^0}\rightarrow f_{CP})}{
                        A(B^0\rightarrow f_{CP})},&\quad& 
  |\rho|^2 \equiv 1 + c_d, 
\end{eqnarray}
and we have neglected the absorptive part of $\langle B^0|{\cal H}^{\Delta B=2}| 
\overline {B^0}\rangle$, which is a good approximation in B meson system. 
The weak phase of the decay amplitude is given 
by $\mbox{arg}[V_{cb}{V_{cs}}^*]$. There is almost no direct CP violation
($c_d=0$) in SM 
because the phase of the penguin amplitude, $b\rightarrow s c\bar c$, is same 
as  $\mbox{arg}[V_{cb}{V_{cs}}^*]$ up to tiny correction. Uncertainty in hadron 
matrix element is cancelled by taking the ratio, so there is no theoretical ambiguity.
We have
\beq
 \frac{Asy[J/\Psi K_s]}{\sin (\Delta M_B t)} = 
 \mbox{Im}\left[  \frac{|M_{12}^B|}{M_{12}^B}
       \frac{V_{cb}{V_{cs}}^*}{{V_{cb}}^*V_{cs}}
       \frac{{V_{cd}}^* V_{cs}}{V_{cd}{V_{cs}}^*} e^{-2i\delta_1}\right]
        = -\sin (\phi_M + 2 \phi_c + 2 \delta_1), \label{phi1}
\eeq
where $\phi_M\equiv \mbox{arg}[M_{12}^B]$, 
$\phi_c \equiv \mbox{arg}[{V_{cb}}^* V_{cd}]$ and 
$\delta_1 \equiv \mbox{arg}[V_{ud}{V_{us}}^*] 
-\mbox{arg}[V_{cd}{V_{cs}}^*] + \pi$ \cite{tk}. 
In SM $\phi_M = - 2  \mbox{arg}[{V_{tb}}^* V_{td}]$, 
so that the righthand-side of eq.(\ref{phi1}) 
becomes $-\sin 2 \phi_1$ up to negligible  
correction of $\delta_1=O(10^{-3})$. The simulation tells us we get the error  
$\delta (\sin 2 \phi_1) = 0.08$ with $10^8$ B mesons. 
\ssect{$\phi_2$}
The $CP$ angle $\tilde\phi_2$ is measured in a similar manner 
as  $\phi_1$ by using $B^0, \overline{B^0} \rightarrow \pi\pi$ decay.
One difference is that there is a penguin contribution which cause direct CP 
violation here. If we can neglect it, we have an asymmetry;
\beq
 \frac{Asy[\pi\pi]}{\sin (\Delta M_B t)} =
  -\mbox{Im}\left[ 
           \frac{|M_{12}^B|}{M_{12}^B} \frac{V_{ub}{V_{ud}}^*}{{V_{ub}}^*V_{ud}}
        \right] = \sin (\phi_M +2\phi_u),
\eeq
where $\phi_u \equiv \mbox{arg}[{V_{ub}}^* V_{ud}]$. The righthand-side of 
the above equation becomes $\sin [2(\pi- \phi_2)]=- \sin 2\phi_2$ in SM.
We will get $\delta (\sin 2 \phi_2) = 0.15$ with $10^8$ B mesons according to 
the simulation. The error will get smaller as statistics increases.
When the penguin contribution is not negligible, we need isospin analysis 
to remove the penguin pollution \cite{iso}. It needs $\pi^0\pi^0$ identification 
which is a challenge for experiment, and the precision gets worse.
We can use $\rho\pi$ mode instead which is easier for experiment. The precision 
given by the simulation is $\delta\phi_2 = 20^\circ$ with $10^8$ B mesons. 
\ssect{$\phi_3$}
The rest of the CP angles $\tilde\phi_3$ is to be measured at B factories
from the decays 
$B^{\pm} \rightarrow \{D^0, \overline{D^0}, D_{CP}\}K^{\pm}$ or 
$B^0 ( \overline{B^0})\rightarrow \{D^0, \overline{D^0}, D_{CP}\}K_S$ \cite{phit}, 
where $D_{CP}$ is a CP eigenstate of neutral D meson which is identified by its 
decay into $K_S\pi^0, K_S\omega, K_S\phi$ or $K^+K^-$. This is a direct CP violation 
process where two amplitudes, $A(b\rightarrow c\bar u s)$ and 
$A(b\rightarrow u\bar c s)$ interfere. The corresponding weak phase is given by 
\beqar
-\arg \left[
    \frac{{V_{cb}}^* V_{us}}{{V_{ub}}^* V_{cs}}
    \frac{V_{cs}{V_{us}}^*}{{V_{cs}}^*V_{us}}
    \right]
&=& -\arg \left[
{(V_{cb}}^*V_{cd})( V_{ub} {V_{ud}}^*)( {V_{us}}^* V_{ud})( {V_{cs}} {V_{cd}}^*)
\right] \nonumber \\
&=& \phi_u - \phi_c -\delta_1 +\pi, 
\eeqar
which becomes $\phi_3$ up to tiny correction of $\delta_1$ in SM.
There is no theoretical ambiguity here. 
The precision according to the simulation is 
$\delta\phi_3 = 25^\circ$ and $15^\circ$ for neutral B and charged B mode, respectively 
with $10^8$ B mesons. (There is one comment here. In the simulation 
the ratio 
$\Gamma (B^- \rightarrow \overline{D^0}K_S)/\Gamma (B^- \rightarrow D^0K_S)$ 
is taken freely from 0.1 to 1.0. But it might be more small, $O(10^{-2})$ \cite{asd}.
Then the precision gets worse.) 
\ssect{Precision with $10^9$ B}
Here we summarize the precision of each measurement. Precision with $10^9$ B 
meson is estimated by naively scaling with $1/\sqrt{\mbox{number}}$.
\begin{center}
\begin{tabular}[c]{c|cccccc}\hline \rule[-3mm]{0mm}{9mm}
& $\frac{\delta (|V_{cb}|)}{|V_{cb}|}$ & $\frac{\delta (|V_{ub}|)}{|V_{ub}|}$ &
  $\frac{\delta (|{V_{tb}}^*V_{td}|)}{|{V_{tb}}^*V_{td}|}$ & 
$\delta(\sin 2\phi_1)$ & 
$\delta (\sin 2\phi_2)$ &$\delta (\phi_3)$ \\
\hline\hline
exp. error ($10^8$ B) &  1 \% & 10 \% &  1 \% & 0.08 & $\ge 0.15$ & $\ge 15^\circ$ \\
exp. error ($10^9$ B) &  $<1$ \% & 3 \% & $<1$ \% & 0.03 & 
                             $\ge 0.05$ & $\ge  5^\circ$ \\ 
theo. error           &  3 \% & 20 \% & 20 \% & 0 & 0 & 0 \\ \hline
\end{tabular}
\vspace*{4mm}

Table 2 : Expected precision of measurements 

\end{center}
Note that this estimation is optimistic in experiment and no theoretical 
development is assumed. The precision of sides is limited by theoretical 
error which is due to uncertainty in hadron effect evaluation, while 
that of angles by experimental error due to small branching ratio and 
difficulty in identification of decay modes. The experimental error 
will be able to get smaller as statistics grows, so the angle measurement 
seems to be promising for the precise determination of the unitarty 
triangle, or KM matrix elements. But there is a possibility of mis-determination 
of the unitarity triangle with angle measurement alone 
if some new physics exists other 
than SM \cite{xsw}. 

Before closing this section one more comment should be given. Unitarity triangle is 
often drawn rescaled by $|{V_{cb}}^*V_{cd}|$. There is about 7 \% error in the 
present value of $|V_{cd}|$ \cite{PDG}. So we need more precise determination 
of $|V_{cd}|$ in the precise determination of triangle by sides 
even if we get precise values of $|V_{cb}|$, $|V_{ub}|$ and 
$|{V_{tb}}^*V_{td}|$. 
\sect{Looking for new physics}
Here we discuss how to check SM and explore new physics as 
the data of unitarity triangle become available and more precise. 
At present we have the data of sides and CP violation in K meson 
mixing. They are consistent with one another in SM. It would be 
difficult to find inconsistency with these data alone without 
great advance in theoretical treatment even if the statistics of 
experimental data gets higher. 

\begin{figure}[hbtp]
  \begin{center}
    \leavevmode
    \epsfig{figure=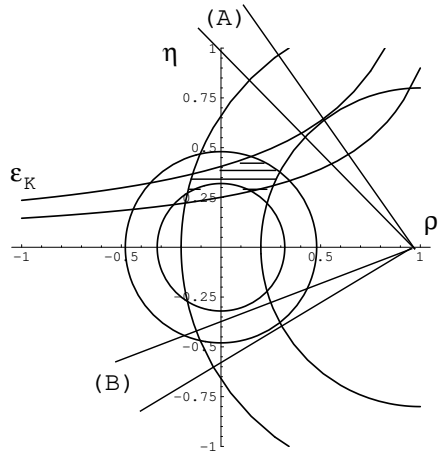,height=6cm}
    \caption{Two cases of inconsistency. The measured 
     $\phi_1$ is too large in the case (A), 
    negative in the case (B).
 }
    \label{fig:newa} 
  \end{center}
\end{figure}

After the B factories begin physics run,  $\phi_1$ will be 
the first to be obtained among the three CP angles. Many of new physics 
can affect $B^0$-$\overline {B^0}$ mixing, so there is a possibility 
of finding inconsistency among $\phi_1$ and other data. New physics is 
likely to appear at the loop level as the new particle is in general 
too heavy to contribute at the tree level. when new physics contributes  
to $M_{12}^B$ with different phase from that in SM, $\phi_1$ deviates 
from the SM value. 
For example, if $\phi_1$ should be proved to be negative, 
it is inconsistent with the CP violation in 
K meson system ($\epsilon_K$). In other case 
it might be found too large to be consistent. 
They are schematically shown in 
Fig.(\ref{fig:newa}). 
The $B^0$-$\overline {B^0}$ mixing matrix elements 
$M_{12}^B$ has almost the same phase with that of SM in 
multi Higgs model with natural flavor conservation 
and minimal SUSY standard model \cite{mbp}, so these models are not likely to 
show up in this stage. While the models with extra quarks \cite{mgen}, 
left-right model \cite{ktw} can  significantly alter the phase of $M_{12}^B$ 
and might be explored.

With more than $10^9$ B meson data the second stage of B physics begins.
All the angles and sides would be obtained then with high precision. 
Also other data of B physics, radiative $b$ decays and so on, would be 
also available then. We can make a systematic study of new physics 
search \cite{grlo}. 

If the data agree with one another very well, SM is confirmed also
in Yukawa sector. Then we can check the hadron physics by comparing the 
experimental value with theoretical value. For example, the hadron 
matrix element of $B^0$-$\overline {B^0}$  mixing, 
$B_Bf_B^2$, can be obtained from experimental value of $\Delta M_B$, $m_t$ 
once KM matrix elements are fixed by other methods. It should be 
compared with the value predicted by lattice QCD. 
There are many works on the quark mass matrices. We can check 
the predictions of KM matrix elements, and discriminate appropriate 
mass matrices, which is benefitable for the explore of higher theory, 
GUTs and strings.   
More severe constraint on new physics is obtained, which gives 
helpful information for the future experimental projects.  
\sect{Concluding remarks}
$10^9$ or more B meson data can give a significant impact on 
the particle physics. When the proposed experimental projects 
on B physics goes well in every respects , we will be 
able to obtain $10^9$ B in 2010 or so. 
Then we can fix KM matrix elements precisely or find a signal of 
new physics beyond SM. This goal can be attained more shortly  
if we get more clear identification of 
decay modes and theoretical development in hadron physics treatment 
are realized. Even in the worst case where only one CP angle ($\phi_1$) 
is obtained, we can find or constrain new physics which 
contributes significantly to $B^0$-$\overline {B^0}$ mixing. 
The second stage of B physics with $10^9$ B will be as promising as 
LEP experiments.

\end{document}